\documentclass[preprint]{aastex}
\usepackage{amsmath}
\usepackage{amssymb}
\usepackage{graphicx}
\usepackage{multicol}
\usepackage{fixltx2e}
\RequirePackageWithOptions{multicol}

\shorttitle{Scatter Broadening at 327 MHz}
\shortauthors{Krishnakumar et al.}
\begin{document}
\def\SPSB#1#2{\rlap{\textsuperscript{{#1}}}\SB{#2}}
\def\SP#1{\textsuperscript{{#1}}}
\def\SB#1{\textsubscript{{#1}}}

\newcommand{\vdag}{(v)^\dagger}
\newcommand{\myemail}{skywalker@galaxy.far.far.away}

\title{Scatter broadening measurements of 124 pulsars at 327 MHz}

\author{M.A. Krishnakumar \altaffilmark{1,2}, D. Mitra \altaffilmark{2}, A. Naidu \altaffilmark{2}, B. C. Joshi \altaffilmark{2} and P. K. Manoharan \altaffilmark{1,2}}
\altaffiltext{1} {Radio Astronomy Centre, NCRA-TIFR, Udagamandalam, India}
\altaffiltext{2} {National Centre for Radio Astrophysics, Tata Institute of Fundamental Research, Pune, India}
\email{kkma@ncra.tifr.res.in}

\begin{abstract}
We present the measurements of scatter broadening time-scales ($\tau_{sc}$) for 
124 pulsars at 327 MHz, using the upgraded Ooty Radio Telescope (ORT). 
These pulsars lie in the dispersion measure range of 37 --- 503 pc cm$^{-3}$ and declination 
($\delta$) range of  $-$57$^{\circ} < \delta< 60^{\circ}$. New $\tau_{sc}$ estimates for 58 
pulsars are presented, increasing the sample of all such measurements by 
about 40\% at 327 MHz. Using all available $\tau_{sc}$ measurements in the literature, we 
investigate the dependence of $\tau_{sc}$ on dispersion measure. Our measurements, together 
with previously reported values for $\tau_{sc}$, affirm that the ionized interstellar medium 
upto 3 kpc is consistent with Kolmogorov spectrum, while it deviates significantly beyond this distance.  
\end{abstract}


\section{Introduction}

Free electrons in the interstellar medium (ISM) affect the pulsar signal in three different
 ways: cold plasma dispersion, free-free absorption and scattering. The broadband pulsar signal 
gets dispersed in the ISM resulting in delay in pulse arrival time, which is a function of wavelength, 
$\lambda$, and its magnitude is described by the quantity known as dispersion 
measure $DM=\int^{D}_{0} n_e dl$ pc cm$^{-3}$, where $n_e$ is the column 
electron density per cm$^{3}$ and D is the distance to the pulsar in parsec. $DM$ for a pulsar is 
obtained as a part of its discovery and if the distance $D$ to the pulsar is known (by parallax 
method for example), then the mean electron density in the ISM can be estimated. 
Alternatively, if the mean $n_e$ is known (either from observations or from the free electron 
density distribution model of the Galaxy such as NE2001 by \citealt{b5}), the approximate distance to the pulsar 
can be determined. The pulsar signals also get absorbed by the free electrons in the ISM. This process 
is known as free-free absorption which is responsible for the turnover observed in many of the 
pulsar spectra at lower frequencies \citep{sie73}.
Pulsar signals are also affected by the fluctuations of the electron density in the ISM. Such 
fluctuations give rise to random variations in the refractive index of the medium. The pulsar 
signal traverses through such irregularities and gets scattered in the process. This phenomenon, 
known as `Interstellar scattering' of pulsar signals, was first investigated by \citet{sch68}. 
It essentially arises due to multipath propagation of the pulsar signal. \citet{w72} demonstrated 
that a narrow pulse with a sharp rise time, while passing through a Gaussian distribution of irregularities 
will broaden in time due to scattering, giving rise to an exponential decay of the pulse with a 
characteristic timescale $\tau_{sc}$, known as the scatter broadening time. The multipath propagation 
also leads to a diffraction pattern in the observer's plane, which decorrelates over a characteristic 
bandwidth $\delta\nu_d$, such that $\tau_{sc} \delta\nu_d = C_1$, where $C_1$ is of the order of unity. 
These scattering parameters are strongly frequency ($\nu$) dependent and scale as $\nu^{-\alpha}$, 
where $\alpha$ is the power law index. 
If the same volume of electrons are responsible for dispersion and scattering, a simple scaling 
relation $\tau_{sc} \propto \nu^{-4} DM^{2}$, can be shown to exist in the model of \citet{sch68}. 
In general, the scattering strength can be attributed to a power law electron density spectrum 
of the form $P_{n_e}(q) = C^{2}_{n_e} q^{-\beta}$, where $q$ is the wavenumber, and $\alpha = 
2\beta/(\beta-2)$ \citep{rick77}. For a Kolmogorov distribution of irregularities, where $\beta 
= 11/3$, the scaling relation is expected to be $\tau_{sc} \propto C^{2}_{n_e}\nu^{-4.4} 
DM^{2.2}$ \citep{rom86}. Several attempts to verify these scaling relations and fathom an 
understanding of the ionized component of the ISM exist in the literature \citep{b13,b12,
lkmll01,b10,b2,b8}. These investigations show that the observed values of $\alpha$ is much lower 
($\le$ 4.0) than the theoretical value of 4.4 for several of the high $DM$ pulsars. Hence, accurate 
$\tau_{sc}$ measurements over multiple frequencies are essential for these studies. 

At meter wavelengths, the $\tau_{sc}$ values for $DM > 100$ pc cm$^{-3}$ dominate over the intrinsic 
pulse width and it is possible to obtain reliable $\tau_{sc}$ measurements. A survey of a number 
of highly dispersed pulsars with a possible detectability at 327 MHz with the aim of finding 
$\tau_{sc}$ was carried out at Ooty Radio Telescope (ORT) by \citet{b13} and \citet{b12} for 47 pulsars 
out of 706 pulsars discovered by that time. Currently $\sim$2200 pulsars are known and the ORT has been 
upgraded for improved sensitivity. Hence, a large number of accurate $\tau_{sc}$ measurements at 
327 MHz using the ORT are now possible. With this aim, we have launched a survey of $\tau_{sc}$ 
measurements. In this paper, we report results from the current survey, which includes several 
new $\tau_{sc}$ measurements and significantly improved $\tau_{sc}$ estimates for pulsars with 
previously measured values.


\section {Observations and Data reduction}

Observations were conducted for 124 pulsars from 2013 January to 2014 September using the ORT, 
situated in southern India at a latitude of 11$^\circ$ N. It has an offset parabolic cylindrical 
reflector, 530-m long in north-south direction and 30-m wide in east-west direction, placed on a 
hillside with a north-south slope of 11$^\circ$ \citep{b17}. 

The feed array consists of 1056 dipoles, oriented along the north-south direction. Hence, the 
telescope is sensitive only to the north-south polarization. The telescope operates at a central 
frequency of 326.5 MHz. The front-end electronics downconverts the signal to an intermediate 
frequency (IF) of 30 MHz with a bandpass of 16 MHz \citep{b15}. The 1056 dipoles are grouped 
in 22 modules with half of them each in north and south halves of the telescope. A separate IF is 
obtained from each half after appropriately phasing the modules and combining their output to form 
12 beams, which cover $36'$ in the sky. 

Recently, a new Pulsar ORT New Digital Efficient Receiver, PONDER, has been commissioned at the ORT to process the 
IF outputs \citep{nbmk14}. The two IF outputs of beam 7 (central beam) are converted to 
baseband and 8-bit digitized in PONDER after processing through a 16-MHz low pass filter. 
The digitized signals from the two halves are added in PONDER and a 4096-point Fast Fourier Transform 
is carried out in real-time across the 16 MHz passband, providing 2048 channels across the passband 
with a spectral resolution of 7.8125 KHz. The channelised data were then incoherently dedispersed in
real-time, to the nominal $DM$ taken from the ATNF pulsar catalogue\footnote{http://www.atnf.csiro.au/people/pulsar/psrcat/} \citep{mht05}. 
PONDER was also used to fold the dedispersed data in real-time to an appropriate number of bins, 
using TEMPO2 predictors\footnote{http://www.atnf.csiro.au/research/pulsar/tempo2/} \citep{hem06} 
with most of the ephemerides taken from ATNF pulsar catalogue. In some cases, particularly for glitching 
pulsars, ephemerides from recently published rotational parameters are used. See \citet{nbmk14} for 
more details on PONDER and the real-time processing capabilities used for this study. The folded profiles 
were then analysed using the procedure described in Section 2.2. 

The ORT has been recently upgraded and currently has a sensitivity which is 1.6 times better than 
what it was in the previous similar studies \citep{b13,b12}. PONDER is designed to digitize the 16 MHz 
bandwidth in contrast with the previous studies, which used a bandwidth of 10 MHz, giving 
a further improvement in sensitivity by a factor of 1.2. Additionally, the telescope has been refurbished 
which improved the gain of the telescope by a factor of 1.3 bringing it to 3.3 K/Jy (eg., \citealt{mano10,nbmk14}).

PONDER has been particularly useful for this project due to the following reasons. Firstly, it eliminated 
the need for the storage of raw data with its real-time capability for channelisation, incoherent 
dedispersion and folding. This results in a smaller dedispersed time series and integrated profile as 
standard data products, which allowed a faster observations to 
analysis turn-around time, even for 8 hour long observations. Secondly, it has a flexible spectral resolution, which enables minimization 
of the dispersion smear across a channel, and consequently allows for robust estimation of scatter-broadening 
parameters. It also has real-time 
coherent dedispersion capability eliminating this smear completely. 
For this project, a fixed spectral resolution of 7.8125 KHz was used. 
Lastly, it will allow real-time 
measurements of dynamic spectra, which will also be useful for 
measurements of ISM effects on pulsed 
signals in future (See \citealt{nbmk14} for more details). 
The above mentioned capabilities are 
useful for future similar investigations of pulsar scatter-broadening.

\subsection{Source Selection}
We selected 124 pulsars for our study using the following method. The pulsars in our sample lie in 
the ORT observable declination, ($\delta$) range of $-57^{\circ}< \delta < 60^{\circ}$ and have a flux 
density at 327 MHz greater than 1 mJy. The flux density at 327 MHz was estimated using available 
spectral index in the ATNF catalogue. Where the spectral index was not available, flux density at the 
frequency available in the ATNF catalogue was scaled using a spectral index of -1.6 \citep{lylg95}. 
The above criteria yielded a large number of pulsars, and the list was further pruned by removing 
pulsars with $DM$s that are not sufficiently high, given that their $\tau_{sc}$ values are likely 
to be significantly smaller 
than the intrinsic pulse profile width. Pulsars with sufficiently large $DM$, where the estimated 
scattered pulse width (obtained by scaling an existing $\tau_{sc}$ measurement at another frequency by 
$\lambda^{4.4}$) exceeds the pulsar period were also removed. The final list of pulsars lie in the $DM$ 
range of 37$-$503 pc cm$^{-3}$ and is listed in Table~\ref{tab1}. The observing time for each pulsar was fixed 
in the following manner. While for every pulsar at least 2000 pulses were acquired to get a stable 
profile, pulsars with low signal to noise ratio were integrated until a minimum peak signal to noise 
ratio of 5 was obtained. Most of the observations were for 30 minutes duration, whereas observations 
were carried out in some special cases up to 8 hours due to the low signal to noise ratio of the pulsar.

\begin{figure*}[h]
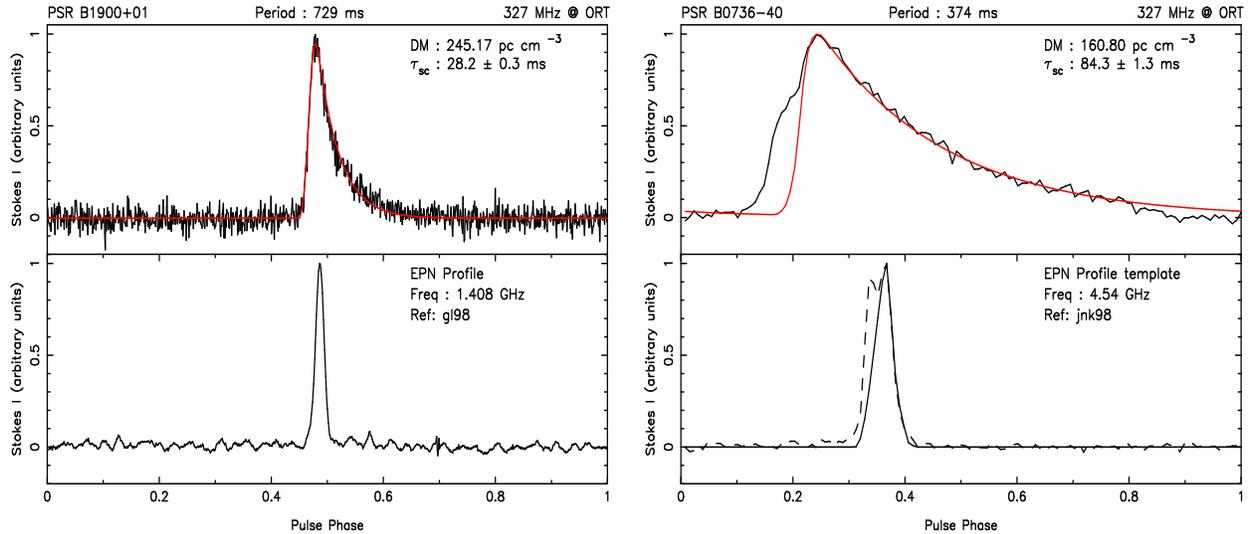

\begin{center}
\begin{tabular}{@{}lr@{}}
{\mbox{\includegraphics[height=80mm,width=70mm,angle=-90.]{B1900+01.ps}}}&
{\mbox{\includegraphics[height=80mm,width=70mm,angle=-90.]{B0736-40.ps}}}\\
\end{tabular}
\caption{The plots shown above are examples of the fitting procedure that has been used to estimate 
$\tau_{sc}$. The solid line in the bottom panel of each plot is the template $P_i(t)$ obtained from an 
existing high frequency observation. The black solid line in the top panel is the observed pulsar 
profile $P(t)$ from ORT, and the red line is the fit to the data obtained using the procedure explained 
in Section~\ref{analp}. The left panel is fit to PSR B1900+01, which is an example where both $P_i(t)$ 
and $P(t)$ have a single component. In the right panel, the example for PSR B0736$-$40 is given. The 
dashed line in the bottom plot is the profile of PSR B0736$-$40 at 4.5 GHz, where two components are 
clearly seen. The solid line, used as the template $P_{i}(t)$, however, is constructed by only retaining 
the trailing component. This template is then used to fit only the trailing component of the observed 
profile as shown by the red line in the top panel. See Section~\ref{analp} for further details.} 
\label{fig1}
\end{center}
\end{figure*}

\begin{figure*}[h]
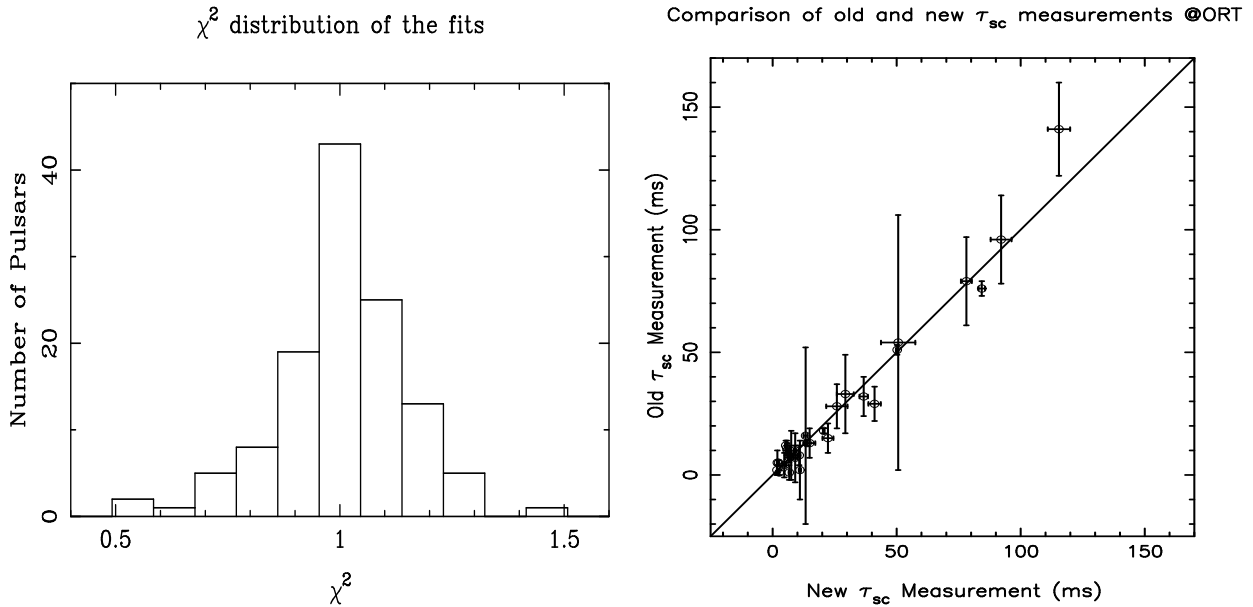

\begin{center}
\begin{tabular}{@{}lr@{}}
{\mbox{\includegraphics[height=80mm,width=80mm,angle=-90.]{chisq.ps}}}&
{\mbox{\includegraphics[height=80mm,width=80mm,angle=-90.]{comparison.ps}}}\\
\end{tabular}
\caption{The plot in the left panel is the histogram of the normalized $\chi^2$ distribution defined by 
equation~\ref{eq2} and that given in Table~\ref{tab1} for all the 124 pulsars observed in the current 
study. The right panel is a comparison of the $\tau_{sc}$ observed in the current study with the previous 
results from ORT \citep{b13,b12}. The new measurements are plotted in the x-axis and the old measurements 
in the y-axis for comparison. A 45$^\circ$ line drawn in black shows that the new measurements and old 
measurements are comparable and are within the error bars.} 
\label{fig2}
\end{center}
\end{figure*}

\subsection{Analysis Procedure}
\label{analp}
The observed pulse profile $P(t)$ is a convolution of intrinsic pulse shape $P\textsubscript{i}(t)$ with 
(1) the impulse response characterizing the scatter broadening in the ISM $s(t)$, (2) the dispersion smear 
across the narrow spectral channel D(t) and the instrumental impulse response, $I(t)$ defined as

\begin{equation}
P(t) = P\textsubscript{i}(t) \ast s(t) \ast D(t) \ast I(t),
\label{eq1}
\end{equation}

\noindent where $\ast$ denotes convolution. The rise times of the receivers and back end are small enough to 
consider the effect of $I(t)$, while $D(t)$ is a rectangular function of temporal width given by the 
dispersion smearing in the narrow spectral channel for incoherent dedispersion. We use the ISM transfer 
function $s(t)=\exp (-t/\tau_{sc})$ \citep{w72}. 

The prime difficulty in extracting $\tau_{sc}$ stems from the fact that the intrinsic unscattered pulse 
profile $P_{i}$ is unknown. In several earlier studies, various assumptions were made regarding $P_{i}$ 
with no clear advantages or disadvantages. For example, \citet{b13} and \citet{b12} assumed the $P_{i}$ 
to have a Gaussian shape and simultaneously fitted for the pulse width and $\tau_{sc}$. \citet{b2} used 
a clean based algorithm, where they claim to recover the intrinsic pulse profile and simultaneously give 
the pulse broadened profile, while \citet{lkmll01} used a high frequency unscattered profile as a model 
for $P_{i}$. In this paper, we follow the method described by \citet{lkmll01}, hereafter referred 
as LKMLL01. For every pulsar in our sample, we searched for a high frequency profile in the EPN data 
base\footnote{http://www.jb.man.ac.uk/research/pulsar/Resources/epn/browser.html}, where the expected 
$\tau_{sc}$ is significantly smaller than the pulse width. In cases, where such profiles were not available, 
often the widths of the profile at a high frequency were found in the literature. We used this width to 
construct a Gaussian profile as a template for $P_{i}$.  Finally, this model $P_{i}$ was used to obtain a 
best fit model by minimizing the normalized $\chi^{2}$ value \citep{lkmll01} defined by 

\begin{equation}
\chi^2 = \frac{1}{(N-4) \sigma\SPSB{2}{off}} \sum_{j=1}^{N}[P\textsubscript{j}(t) - 
P\textsubscript{m}(a,b,c,\tau_{sc})]^2, 
\label{eq2}
\end{equation}

\noindent where $\sigma$\SPSB{2}{off} is the off-pulse root-mean-square (rms), $P\textsubscript{j}(t)$ is the observed 
pulse profile, $P\textsubscript{m}(t)$ is the model profile and $N$ is the total number of bins in the 
profile. The model profile $P\textsubscript{m}$ is scaled with the pulse amplitude $a$, shifted by a constant 
offset $b$ in phase and fitted to a baseline $c$ to minimize the $\chi^2$. For fitting purpose, we use the 
nonlinear fitting routine ``mrqmin'' given in Numerical Recipes \citep{press01}, where the errors in $\tau_{sc}$ 
is obtained from the covariance matrix.

Two examples\footnote{Results of the fitting procedure for all the 124 pulsars and integrated profiles are 
available online at http://rac.ncra.tifr.res.in/data/pulsar/supplementary-material.pdf. All the pulsar profiles used in this study is publicly available in SIGPROC ASCII format which can be downloaded from http://rac.ncra.tifr.res.in/data/pulsar/124-ascii-profiles.tar.} of the fitting 
procedure used to estimate $\tau_{sc}$ values are shown in Figure~\ref{fig1}. The top panel of the left hand 
figure is the fit to the profile of PSR B1900+01, and the bottom panel is the high frequency profile obtained 
at 1.4 GHz, which has a single component and is used as a template $P_i(t)$. The fitting process yields an 
excellent fit to the data (shown in the top panel of the figure) with $\chi^2 \sim 1$ and $\tau_{sc} 
\sim 28.2$ ms. It is important to note that we have ignored the effect of pulse width and component evolution 
with observing frequency in our fitting algorithm. It is known that pulse width increases with increasing 
wavelength as $\lambda^{0.3}$ due to radius to frequency mapping \citep{thor91,mr02}, and hence $\tau_{sc}$ 
value that we obtain is an upper limit. However, in several cases, particularly for high $DM$ pulsars, the 
$\tau_{sc}$ evolution is significantly higher than the width evolution and hence the measured values of 
$\tau_{sc}$ are robust. In the example of PSR B1900+01, the pulse width is expected to increase from a higher 
frequency profile by a factor of 1.5 whereas the expected increase in $\tau_{sc}$ is a factor of 40, indicating 
that the $\tau_{sc}$ measurement is reliable. However, the values of $\tau_{sc}$ can be affected by the 
presence of an unidentified pulse component on the trailing part of the profile, as discussed by LKMLL01.
 They suggested that an increased error bar to 3 standard deviations, is usually sufficient to account for 
these effects. It is noteworthy that for $\sim$70\% of the pulsars in our sample, the above procedure serves 
as an acceptable model.

The remaining $\sim$30\% of pulsars in our sample has multiple component profiles, both at higher and 
lower frequencies (except three cases that are discussed below). For these pulsars, 
applying the method of LKMLL01 is not possible. The profile evolution in pulsars is such that the location and 
amplitude of the high frequency profile components change significantly at a lower frequency. However, if the 
separation of a profile component (identified at both higher and lower frequencies) from the adjacent one is 
significantly larger than $\tau_{sc}$, then the LKMLL01 method can be applied to the component. The second example 
of PSR B0736$-$40 shown in the right hand panel of Figure~\ref{fig1} pertains to such a case.
 The high frequency template for PSR 
B0736$-$40 has two components, and as shown in the figure, we have considered only the trailing component as a 
model and fitted this to the trailing component of the pulse profile at 327 MHz. This has resulted in significantly 
better fitting and hence estimation of $\tau_{sc}$.  We have applied this procedure to our data wherever applicable 
($\sim$30\% of the pulsars presented here). To gain additional confidence in this method, we have performed 
extensive simulations\footnote{Detailed plots and description of these fits are available in the online supplementary 
material.}, where we created  unscattered templates with two Gaussian components and convolved with the exponential 
transfer function to get the scattered profile for an input $\tau_{sc}$ value. We then used the procedure described 
above to extract $\tau_{sc}$ from the scattered profile. Multiple templates were obtained by varying the amplitude 
ratio and the separation of the Gaussian components, and for each case the corresponding $\tau_{sc}$ value was 
estimated. We find that for majority of the cases, where components could be resolved, we are able to recover the 
$\tau_{sc}$ value (within 3 standard deviation). 
When the separation 
between the components and/or the ratio of the amplitudes are too small, 
we tend to overestimate $\tau_{sc}$ by about 10\%, as can be seen in Figure 64 of the supplementary material.

There are 3 pulsars in our sample, namely B1737$-$30, B1834$-$10 and B1859+03, which have a slow rising edge and 
heavily scatter-broadened trailing edge. For these pulsars, the LKMLL01 model is found inadequate, since the 
rising edge of the profiles are poorly modelled.
If we {\em assume} that there is no unidentified emission component that changes the profile shape in the rising 
edge for these pulsars, then there exist a possibility that the transfer function of the ISM towards these pulsars are 
responsible for these profile changes. We have explored whether the transfer function for a uniformly distributed ISM with 
Gaussian density fluctuation (see \citealt{w72,w73,b10,b2}) given by 
$s(t) = \left( \pi^{5} \tau_{sc}^{3}/8t^5\right)^{1/2} exp(-\pi^2 \tau_{sc}/4t)$, can model the data, since the 
rising edge of this transfer function increases much more slowly as seen in the data\footnote{This function is also 
refered as the pulse broadening function PBF2 in \citet{b10,b2}.}. In all the three cases, we found reasonable fits 
and they yielded $\tau_{sc}$ values of 62$\pm$3, 203$\pm$23 and 103$\pm$2, respectively (the fitted profiles are shown 
separately in the online supplementary material). However, before we conclude that the effect seen in these pulsars are 
related to a different transfer function, futher modelling of the evolution of pulse profiles based on high quality 
multifrequency observations need to be done, which is beyond the scope of the present study.

Table~\ref{tab1} lists the estimated $\tau_{sc}$, the error $\delta\tau_{sc}$ and the $\chi^{2}$ for all the 124 
pulsars in our sample. The left panel of Figure~\ref{fig2} shows the distribution of reduced $\chi^{2}$ for our 
sample, which lie in a range of 0.5 to 1.5 and peaks around unity. This confirms that the model profile $P_m$ 
is a good fit to the data for majority of pulsars in our sample. In the right panel of Figure~\ref{fig2}, 
$\tau_{sc}$ values for 31 pulsars which are common between our sample (the abscissa) and that of \citet{b13} and 
\citet{b12} (the ordinate) are shown for comparison. The $\tau_{sc}$ values obtained between these two methods 
are generally in good agreement with each other.

\section{Results and Discussions}
\label{result}

The primary goal of this work is to provide a large database of $\tau_{sc}$ measured at 327 MHz (P-band). In this 
paper, we have estimated $\tau_{sc}$ values for 124 pulsars, in which, as far as we know, 58 are new measurements 
at P-band, which thereby increases the $\tau_{sc}$ measurements at 327 MHz by almost 40\%. Including our current sample 
and measurements available in the literature, we found 154 pulsars for which scattering measurements exist at P-band: 
out of which 136 have $\tau_{sc}$ measurements, with 124 from this paper and 12 from \citet{b13} and \citet{b12}. 18 
have decorrelation bandwidth $\delta\nu_d$ measurements from \citet{bpg99}. Additionally, we found 134 pulsars having 
$\tau_{sc}$ values and 70 pulsars with $\delta\nu_d$ values measured at different frequencies. Thus, currently 
scattering measurements exist for 358 pulsars spanning a DM range of 2.9 -- 1073 pc cm$^{-3}$.

\begin{figure*}[ht]
\centering
\includegraphics[height=150mm,width=150mm,angle=-90]{dm_vs_tau.eps}\\
\caption{The top panel shows the plot of $\tau_{sc}$ versus $DM$ at 327 MHz. Filled circles are $\tau_{sc}$ 
measurements at 327 MHz (these include all the measurements presented in this paper and 12 measurements from 
\citep{b13,b12}). The black filled triangles correspond to the $\delta\nu_d$ measurements from ORT by \citet{bpg99} 
at 327 MHz. The open circles and open triangles correspond to $\tau_{sc}$ and $\delta\nu_d$ measurements respectively
available at a different frequency \citep{ra82,b6,b1,b4,dcw88,joh90,b20,clj92,b2,b8} and scaled appropriately to 327 
MHz (see text for more details). The continuous curve is a least squares fit to the whole data set. Bottom panel shows 
the fractional residual (i.e., (data - model)/model) plotted against $DM$. The fractional residuals are plotted for a 
limited y-range, since there are some anomalous pulsars whose residuals are significantly high.}
\label{taudm}
\end{figure*}

\begin{figure*}[ht]
\begin{center}
\begin{tabular}{@{}c@{}}
{\mbox{\includegraphics[height=80mm,width=160mm]{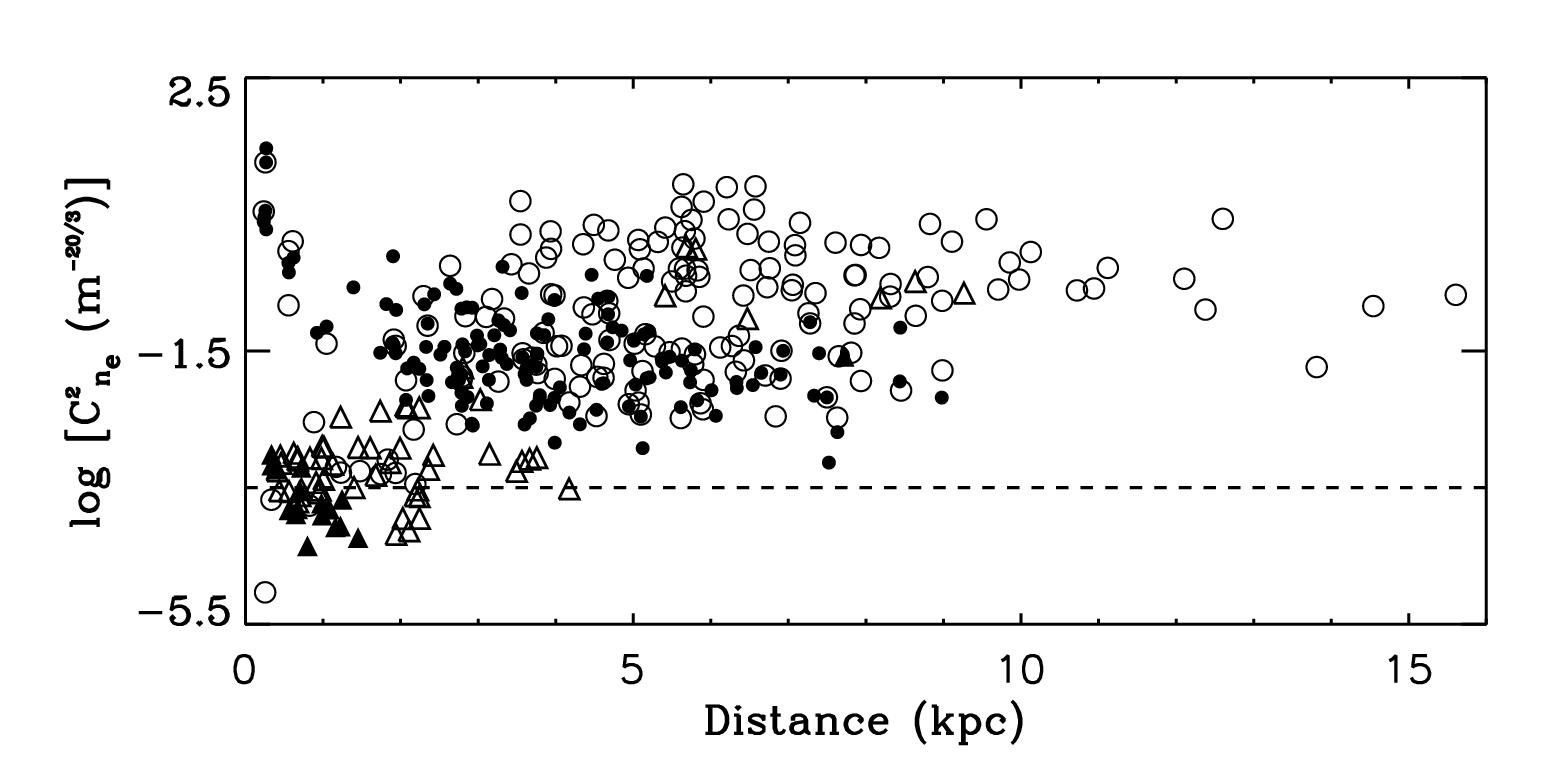}}}\\
{\mbox{\includegraphics[height=80mm,width=160mm]{{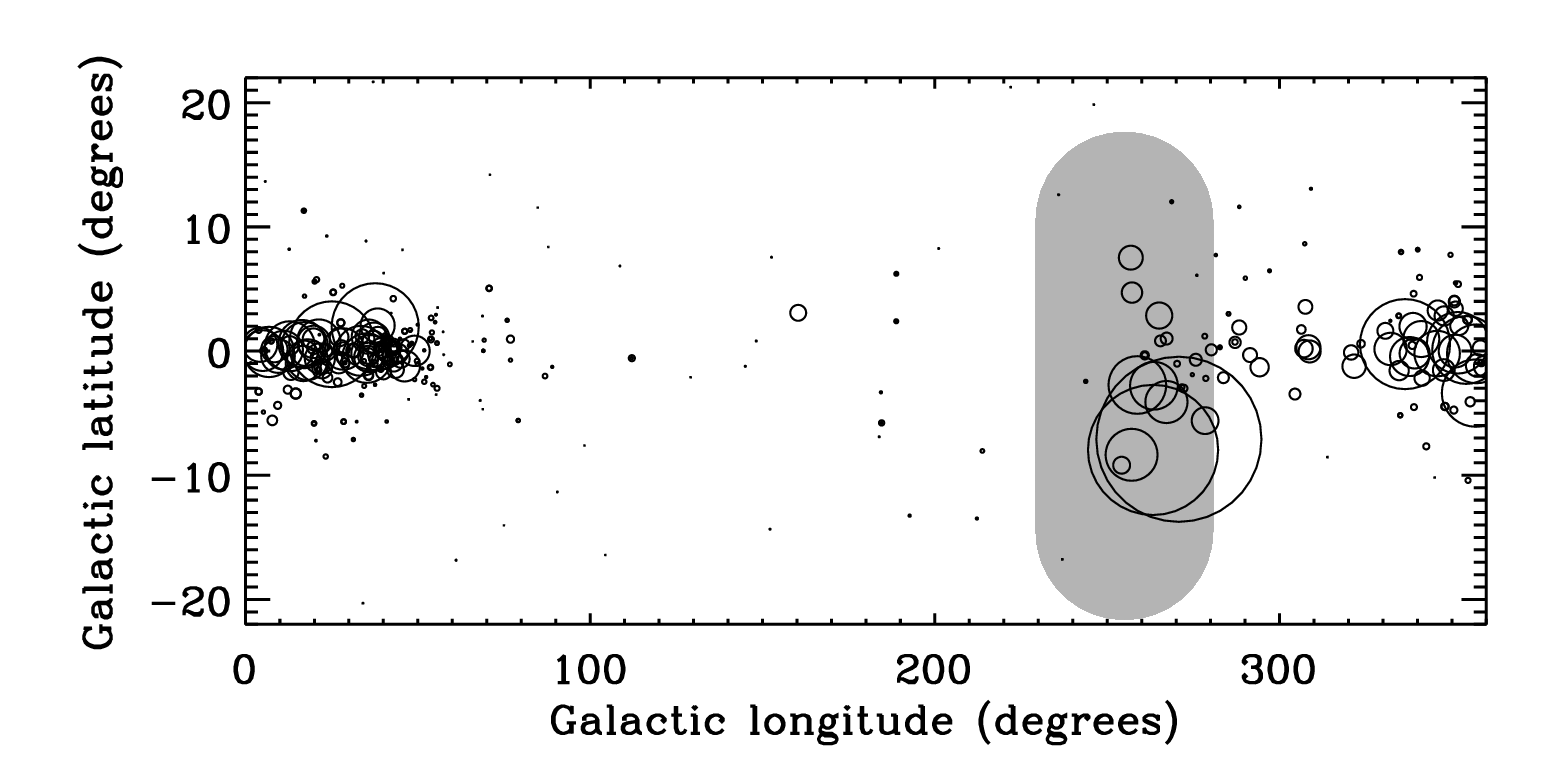}}}}\\
\end{tabular}
\caption{The top panel shows the plot of scattering strength, $log (C_{n_e}^{2})$ against distance to the 
pulsars. The dashed line corresponds to $-$3.5, the canonical value for a homogeneous Kolmogorov medium. The symbols 
in the plot follow the same coding as in Figure~\ref{taudm}. Bottom panel shows the plot of $C_{n_e}^{2}$ of the
pulsars in the galactic plane. The diameter of the circle denote the magnitude of $C_{n_e}^{2}$. This plot 
includes all our measurements and those quoted in the caption of Figure~\ref{taudm}. The shaded area is the Gum 
nebula region.}
\label{cn2}
\end{center}
\end{figure*}

Generally, scattering properties along various lines of sight to the pulsars are examined by studying the relation 
between $\tau_{sc}$ and $DM$. Figure~\ref{taudm} shows the $\tau_{sc}$ versus $DM$ plot at 327 MHz for all the 358 
pulsars. The filled circles are direct measurements at 327 MHz. The black triangles are $\delta\nu_d$ measurements 
at 327 MHz which have been converted to $\tau_{sc}$ using the relation $\tau_{sc} = C_1 / \delta\nu_d$, where $C_1$ 
is assumed to be 1.16 (see \citealt{b10} for discussion). For convenience of discussion, let us define four different 
$DM$ regions in the  $\tau_{sc}$ versus $DM$ plot :  region $R1$ for $DM$ range 0 to 40 pc cm$^{-3}$, region $R2$ for $DM$ 
range 40 to 100 pc cm$^{-3}$, region $R3$ for $DM$ range 100 to 500 pc cm$^{-3}$ and region $R4$ for $DM$ range above 500 pc cm$^{-3}$. 
It may be noted that in the region $R2$, no reliable $\tau_{sc}$ or $\delta\nu_d$ measurements are currently available 
at P-band. In this range at P-band, the $\tau_{sc}$ becomes unreliable since the expected $\tau_{sc}$ values are 
significantly smaller than the profile widths. The $\delta\nu_d$ values are also difficult to measure since they 
decorrelate over a significantly small bandwidth. Further $\tau_{sc}$ measurements at P-band in the region $R4$ is 
not possible since the scattering time exceeds the pulsar period. However, the 327 MHz data reveal the basic feature 
of the $\tau_{sc}$ versus $DM$ relation, where at low $DM$ i.e., region $R1$, the slope is flatter compared to the high 
$DM$ region $R3$ \citep{sut71,rick77,b6,b13}. 

In order to fill in data at P-band for the missing $DM$ ranges $R2$ and $R4$ in Figure~\ref{taudm}, we rely on scattering 
measurements at higher frequencies. In region $R2$, $\delta\nu_d$ measurements and in region $R4$, $\tau_{sc}$ 
measurements are available at higher frequencies. These measurements therefore need to be scaled to 327 MHz using a 
frequency scaling law $\tau_{sc} \propto \nu^{-\alpha}$. The properties of the Kolmogorov spectrum of electron density 
irregularities, which has been widely used as a reference  to understand the $\tau_{sc}$ versus $DM$ curve, predicts 
$\alpha=$4.4. However, observationally there is evidence that the frequency scaling of $\alpha=$4.4 might not be uniform 
across the pulsar sample. For DMs below about 100 pc cm$^{-3}$ \citet{b6} and \citet{joh98} found $\alpha$ to be consistent 
with 4.4, whereas for higher DM values, \citet{lkmll01} and \citet{b8} reported a much smaller value of $\alpha$ $\sim$ 
3.44. The scaling relation for these pulsars were obtained on a case by case basis using detailed modelling and 
measurement of $\tau_{sc}$ and $\delta\nu_d$. \citet{b2} conducted an exhaustive study for obtaining $\alpha$ treating 
it as a parameter (constant across the whole DM range) and fitted for $\alpha$ using direct estimates of $\tau_{sc}$ 
and $\delta\nu_d$ at the observing frequencies. It yielded a value of 3.86$\pm$0.16. Currently more accurate 
measurements are required to establish the frequency scaling as the function of DM. Here we use $\alpha=$ 4.4, which can 
be referenced to the Kolmogorov spectrum and scaled the scattering measurements to 327 MHz using the frequency 
dependence of $\nu^{-4.4}$ (note that the $\delta\nu_d$ measurements are converted to $\tau_{sc}$ using $C_1 = 1.16$). 
The scaled $\delta\nu_d$ and $\tau_{sc}$ values are shown as open triangles in Figure~\ref{taudm}.

The combined $\tau_{sc}$ versus $DM$ plot is obtained using all 
currently available scattering measurement data of 358 pulsars. We fitted the empirical relation $\tau_{sc} 
= A DM^{\gamma} (1 + B DM^{\zeta})$, as was suggested by \citet{b13}. By fixing $\gamma = 2.2$, the term $(1 + B 
DM^{\zeta})$ can provide a useful description for assessing the deviation of $\tau_{sc}$ with $DM$ from the Kolmogorov 
theory. The best form of the fit (shown as solid curve in the figure) corresponds to $\tau_{sc} \rm{(ms)} = 3.6 \times 
10^{-6} DM^{2.2} (1 + 1.94 \times 10^{-3} DM^{2.0})$ which is consistent with earlier studies of \citet{b13} and \citet{b10}. 
Note that there is a large spread in $\tau_{sc}$ for a given $DM$ and in the bottom panel of Figure~\ref{taudm}, the 
fractional residual (i.e. (data - model)/model) is shown, where it appears that with increasing $DM$, the spread in $
\tau_{sc}$ increases. There are a few reasons for the spread in $\tau_{sc}$ measurements. As discussed earlier, the 
methods applied for estimating $\tau_{sc}$ can lead to uncertainties due to ambiguity in the intrinsic pulse shape. 
This mostly affects $\tau_{sc}$ values in regions $R1$ and $R2$ at P-band. However, for larger $DM$s, i.e., regions $R3$ and 
$R4$, $\tau_{sc}$ values are significantly larger than the pulse profile widths, and hence the ISM properties are mostly 
the reason for the observed spread in $\tau_{sc}$. Clearly our analysis reveals several anomalous lines of 
sights, for example PSR B1015$-$56 has a $DM$ of 436 pc cm$^{-3}$ while the measured $\tau_{sc}$ is 13 ms which is a factor 
of 10 lower than the mean $\tau_{sc}$ for that $DM$. Anomalous scattering is a commonly observed phenomenon, and is most 
probably related to inhomogeneous distribution of scattering material along pulsar lines of sight. 

The indirect (i.e. non P-band measured) $\tau_{sc}$ values in Figure~\ref{taudm} depends on $C_1$ and the frequency 
scaling. The change in $C_1$ in the lower $DM$ range can lead to a change in the slope of the $\tau_{sc}$ versus $DM$ 
relation. As argued by \citet{rick77} and \citet{b10}, there is very little allowance in the theory for changing 
$C_1$ values from close to unity. However the frequency scaling of $\nu^{-4.4}$ used in Figure~\ref{taudm} has been 
a subject of scrutiny in recent studies by \citep{lkmll01,b10,b8}. These careful studies of multifrequency $\tau_{sc}$ 
measurements revealed that at high $DM$ ($\gtrsim$500 pc cm$^{-3}$) the frequency scaling is about 3.4, which is significantly 
smaller than the Kolmogorov value. This can result in flattening of the slope in the region $R4$ of the $\tau_{sc}$ 
versus $DM$ curve, and only a systematic multifrequency study to obtain scattering estimates for a large sample of 
pulsars can resolve some of these issues. However there is significant overlap of pulsars with direct P-band 
$\tau_{sc}$ measurements (filled circles) and frequency scaled $\tau_{sc}$ measurements in the $DM$ region $R3$, where no
differences in $\tau_{sc}$ values can be seen. We can hence speculate that for pulsars below $DM$ 500 pc cm$^{-3}$, the 
frequency scaling of $\nu^{-4.4}$ is a good approximation (also a result supported by \citealt{lkmll01,b10,b8}). 

It has been also speculated that the breakdown of the homogeneity condition in the ISM \citep{rick77} can lead to the 
steepening of $\tau_{sc}$ values. This can be checked by evaluating the scattering strength $C_{n_e}^{2}$ along 
pulsar lines of sight. In the top panel of Figure~\ref{cn2}, the estimated $log(C_{n_e}^{2})$ for all the 358 pulsars 
as a function of pulsar distance in kpc is shown. The $C_{n_e}^{2}$ values were obtained for a homogeneous medium 
with Kolmogorov spectrum using the expression given by \citet{b4} as $C_{n_e}^{2} = 0.002 \nu^{11/3} D^{-11/6} \delta
\nu_d^{-5/6}$, where $\nu$ is the observing frequency in GHz, $D$ is distance to the pulsar in kpc and $\delta\nu_d$ 
is the decorrelation bandwidth in MHz. The distance to the pulsars were obtained using the \citet{b5} model of 
Galactic free electron density distribution. The dashed line in the top panel of Figure~\ref{cn2} correspond to the 
case of homogeneous medium where $log(C_{n_e}^{2}) = -3.5$ \citep{joh98}. Our results are consistent with all earlier 
studies \citep{b4,joh98,b10} where lines of sight to pulsars below 3 kpc has a value closer to $-$3.5. For the distant 
pulsars, the deviation is significant towards larger $log(C_{n_e}^{2})$ values which indicate that high level of 
inhomogeneity exists along these lines of sight. The bottom panel of Figure~\ref{cn2} shows the distribution of all 
358 pulsars in the Galaxy, where majority of the pulsars lie towards the inner regions of the Galaxy and hence are 
likely to encounter several HII regions. Note that the enhanced scattering seen around Galactic longitude of 260$^
{\circ}$ is due to the Gum Nebula \citep{b12}.

\section{Summary}
In this paper, we report new observations of 124 pulsars using the Ooty Radio Telescope at 327 MHz with the aim of 
estimating scatter broadening ($\tau_{sc}$) values. Our attempt has yielded 58 new $\tau_{sc}$ measurements at 327 
MHz, which increases the sample with known $\tau_{sc}$ measurement at 327 MHz by about 40\%. The major difficulty 
in estimating $\tau_{sc}$ lies in the uncertainty of knowing the intrinsic (unscattered) pulse shape. We have hence 
used the method suggested by \citep{lkmll01,b10} in which the intrinsic pulse profile is obtained from higher 
frequencies where the profile is unscattered. This method significantly reduces the uncertainty in $\tau_{sc}$ 
measurements. Further we have also assimilated 234 scattering measurements from the literature and have used them to 
study the ISM properties. Our results are consistent with earlier works where, for low $DM$ (or below 3 kpc), the ISM 
is consistent with Kolmogorov spectrum, but it deviates significantly for high $DM$ pulsars. In Figure~\ref{taudm}, the 
$\tau_{sc}$ values at P-band for several pulsars were obtained estimating $\tau_{sc}$ from $\delta\nu_d$ using a 
frequency scaling of $\nu^{-4.4}$ from higher frequencies. The high sensitivity of the upgraded ORT and the 
flexibility of PONDER will be useful to enhance the sample of such observations further, particularly to weaker 
pulsars, which were not covered in this study. We also note that these measurements can be improved by observations of 
pulsars at multiple frequencies with instruments, such as the GMRT, covering the low $DM$ sample with frequencies 
including and below 300 MHz and the high $DM$ sample at 610 or 1420 MHz and such observations are planned in future.

\section{Acknowledgments}

We acknowledge the kind help and support provided by the members of Radio Astronomy Centre during these observations. 
The ORT is operated and maintained at the Radio Astronomy Centre by the National Centre for Radio Astrophysics of the 
Tata Institute of Fundamental Research. We also thank the anonymous referee for a critical review which helped in improving 
the manuscript significantly.

\clearpage

\begin{deluxetable}{clccccccc|cc|c}

\tablecolumns{11}
\tablewidth{0pt}  
\tabletypesize{\small}
\tablecaption{Results. For each pulsar, the table lists its position in galactic longitude and latitude, dispersion measure, measured $\tau_{sc}$ with error, 3$\sigma$ pulse width, normalized $\chi^2$, number of degrees of freedom, distance to the pulsar and expected $\tau_{sc}$ taken 
from \citet{b5} model and the scattering strength, $C_{n_e}^{2}$ from our measurements. Pulsars shown in bold 
characters are the ones observed with ORT by \citet{b13} and \citet{b12}.}

\tablehead{
\colhead{No.}  & \colhead{PSR} &  \colhead{l} & \colhead{b} & \colhead{DM} & \colhead{{$\tau_{sc}$}} & \colhead{W\SB{3$\sigma$}} & \colhead{$\chi^2$} & \colhead{N\textsubscript{dof}} & \colhead{D} & \colhead{{$\tau$\SPSB{NE01}{sc}}}  &  \colhead{log $C_{n_e}^{2}$ } \\

\colhead{} & \colhead{} &  \colhead{(deg)} & \colhead{(deg)} & \colhead{pc cm$^{-3}$} & \colhead{(ms)} & \colhead{(ms)} & 
\colhead{} & \colhead{} & \colhead{(kpc)} & \colhead{(ms)} & \colhead{($m^{-20/3}$)}\\
}
\startdata
1 & B0447-12 $\dagger$ & 211.08 & -32.63& 37.04 & 4.0$\pm$0.1 & 33.77 & 1.0 & 464 & 1.9 & 0.02 & -1.38\\
2 & B0450-18 $\dagger$ & 217.08 & -34.09& 39.90 & 0.75$\pm$0.01 & 37.33 & 0.8 & 487 & 2.4 & 0.02 & -2.16\\
3 & B0458+46 & 160.36 & 3.08& 42.19 & 19$\pm$1& 36.79 & 1.2 & 483 & 1.4 & 0.05 & -0.57\\
4 & B0523+11 $\dagger$ & 192.70 & -13.25& 79.34 & 1.02$\pm$0.04& 19.90 & 1.0 & 478 & 3.1 & 0.21 & -2.27\\
5 & B0531+21 & 184.56 & -5.78& 56.79 & 1.99$\pm$0.02& 27.90 & 1.0 & 124 & 1.7 & 0.28 & -1.52\\
6 & B0559-05 $\dagger$ & 212.20 & -13.48& 80.54 & 1.6$\pm$0.1& 26.52 & 1.0 & 415 & 3.9 & 0.43 & -2.29\\
7 & B0611+22 & 188.79 & 2.40& 96.91 & 1.74$\pm$0.03& 19.51 & 0.8 & 392 & 2.1 & 0.3 & -1.75\\
8 & B0621-04 $\dagger$ & 213.79 & -8.04& 70.83 & 2.6$\pm$0.3& 67.47 & 1.1 & 382 & 2.8 & 0.2 & -1.84\\
9 & B0626+24 $\dagger$ & 188.82 & 6.22& 84.19 & 2.01$\pm$0.04& 28.63 & 1.0 & 482 & 2.2 & 0.32 & -1.76\\
10 & {\bf B0736-40} $\dagger$ & 254.19 & -9.19& 160.80 & 84$\pm$1& 254.84 & 1.0 & 125 & 2.6 & 117 & -0.51\\
11 & B0740-28 $\dagger$ & 243.77 & -2.44& 73.78 & 0.71$\pm$0.01& 17.28 & 1.1 & 383 & 2.1 & 0.0001 & -2.22\\
12 & J0749-4247 & 257.07 & -8.35& 104.59 & 3.4$\pm$0.4& 29.95 & 0.8 & 253 & 0.3 & 21 & 0.46\\
13 & {\bf B0808-47} & 263.30 & -7.96& 228.30 & 78$\pm$2& 134.67 & 1.2 & 253 & 0.3 & 131 & 1.26\\
14 & B0818-41 $\dagger$ & 258.75 & -2.74& 113.40 & 9.7$\pm$0.1& 234.38 & 1.1 & 124 & 0.3 & 24 & 0.55\\
15 & B0833-45 & 263.55 & -2.79& 67.99 & 5.41$\pm$0.03& 50.97 & 0.9 & 125 & 0.2 & 7.01 & 0.39\\
16 & {\bf B0835-41} & 260.90 & -0.34& 147.29 & 1.99$\pm$0.01& 32.30 & 0.5 & 1021 & 1.0 & 2.9 & -1.14\\
17 & B0840-48 & 267.18 & -4.10& 196.85 & 5$\pm$1& 20.14 & 1.0 & 253 & 0.3 & 93 & 0.28\\
18 & {\bf B0844-35} $\dagger$ & 257.19 & 4.71& 94.16 & 5$\pm$1& 82.48 & 0.9 & 430 & 0.6 & 0.00001 & -0.35\\
19 & {\bf B0853-33} & 256.85 & 7.517& 86.64 & 6.7$\pm$0.3& 34.15 & 1.0 & 331 & 0.6 & 0.00001 & -0.21\\
20 & J0857-4424 & 265.46 & 0.82& 184.43 & 16$\pm$3& 26.81 & 1.0 & 252 & 1.9 & 11 & -0.90\\
21 & {\bf B0903-42} & 265.08 & 2.86& 145.80 & 11$\pm$1& 45.24 & 1.3 & 253 & 0.6 & 7 & -0.13\\
22 & J0905-4536 $\dagger$ & 267.24 & 1.01& 179.70 & 18$\pm$2& 216.19 & 1.1 & 253 & 2.3 & 0.02 & -0.82\\
23 & B0906-49 & 270.27 & -1.02& 180.37 & 7$\pm$1& 65.13 & 1.0 & 97 & 2.6 & 0.03 & -1.44\\
24 & {\bf B0950-38} & 268.70 & 12.03& 167.00 & 11.0$\pm$0.4& 96.60 & 1.1 & 253 & 6.5 & 0.16 & -1.99\\
25 & {\bf B1015-56} & 282.73 & 0.34& 439.10 & 13$\pm$1& 47.20 & 2.4 & 126 & 8.9 & 112 & -2.18\\
26 & J1036-4926 $\dagger$ & 281.52 & 7.73& 136.53 & 2$\pm$1& 11.96 & 1.1 & 125 & 4.0 & 1.2 & -2.19\\
27 & {\bf B1039-55} $\dagger$ & 285.19 & 2.99& 306.50 & 9$\pm$1& 45.73 & 1.1 & 254 & 6.3 & 34 & -2.04\\
28 & B1044-57 & 287.07 & 0.73& 240.20 & 19$\pm$1& 37.52 & 1.2 & 125 & 4.4 & 24 & -1.47\\
29 & B1119-54 & 290.08 & 5.87& 204.70 & 8.6$\pm$0.2& 58.60 & 1.2 & 125 & 5.2 & 8.3 & -1.90\\
30 & J1123-4844 $\dagger$ & 288.29 & 11.61& 92.92 & 1.1$\pm$0.2& 12.43 & 1.1 & 253 & 2.9 & 0.2 & -2.17\\
31 & J1210-5559 $\dagger$ & 297.14 & 6.42& 174.35 & 0.9$\pm$0.2& 6.56 & 1.0 & 253 & 4.3 & 7 & -2.57\\
32 & B1317-53 & 307.30 & 8.64& 97.60 & 1.4$\pm$0.2& 10.93 & 0.9 & 125 & 2.3 & 1.4 & -1.92\\
33 & B1325-49 $\dagger$ & 309.12 & 13.07& 118.00 & 0.8$\pm$0.4& 63.54 & 1.0 & 509 & 3.7 & 1.1 & -2.30\\
34 & B1523-55 & 323.64 & 0.59& 362.70 & 59$\pm$6& 49.16 & 1.0 & 253 & 5.2 & 128 & -1.21\\
35 & {\bf J1557-4258} & 335.27 & 7.95& 144.50 & 5.3$\pm$0.3& 22.01 & 1.1 & 253 & 4.6 & 10 & -1.98\\
36 & {\bf B1600-49} $\dagger$ & 332.15 & 2.44& 140.80 & 1.77$\pm$0.04& 19.18 & 0.6 & 253 & 5.1 & 15 & -2.46\\
37 & {\bf B1609-47} & 334.57 & 2.84& 161.20 & 7$\pm$1& 22.41 & 0.9 & 253 & 3.7 & 9.4 & -1.72\\
38 & {\bf J1614-3937} $\dagger$ & 340.00 & 8.21& 152.44 & 4.4$\pm$0.3& 28.64 & 0.9 & 254 & 3.8 & 1.5 & -2.20\\
39 & B1620-42 & 338.89 & 4.62& 295.00 & 52$\pm$6& 27.06 & 1.0 & 253 & 6.6 & 13 & -1.44\\
40 & J1625-4048 $\dagger$ & 340.61 & 5.93& 145.00 & 4$\pm$2& 147.22 & 1.1 & 252 & 3.1 & 5 & -1.56\\
41 & B1635-45 & 338.50 & 0.46& 258.91 & 26$\pm$1& 63.79 & 1.1 & 362 & 3.8 & 60 & -1.24\\
42 & J1648-3256 & 349.59 & 7.75& 128.28 & 4.4$\pm$0.3& 16.23 & 1.1 & 396 & 3.1 & 2.7 & -1.72\\
43 & {\bf B1647-52} & 335.01 & -5.17& 179.10 & 6.7$\pm$0.4& 29.77 & 1.1 & 253 & 3.8 & 12 & -1.73\\
44 & J1700-3312 $\dagger$ & 351.06 & 5.49& 166.97 & 4$\pm$1& 57.73 & 1.0 & 397 & 3.6 & 3 & -1.92\\
45 & B1658-37 & 347.76 & 2.83& 303.40 & 543$\pm$55& 1112.15 & 1.1 & 125 & 5.2 & 61 & -0.40\\
46 & B1700-32 $\dagger$ & 351.79 & 5.39& 110.31 & 4.7$\pm$0.1& 71.10 & 0.9 & 406 & 2.3 & 2.5 & -1.44\\
47 & J1703-4851 & 338.99 & -4.51& 150.29 & 10$\pm$1& 60.01 & 0.9 & 253 & 3.0 & 2.9 & -1.42\\
48 & {\bf J1705-3423} & 350.72 & 3.98& 146.36 & 41$\pm$3& 68.84 & 0.9 & 252 & 2.9 & 10 & -0.86\\
49 & J1708-3426 & 351.08 & 3.41& 190.70 & 120$\pm$13& 51.37 & 1.0 & 125 & 3.6 & 11 & -0.65\\
50 & {\bf B1718-32} & 354.56 & 2.53& 126.06 & 13.9$\pm$0.2& 51.84 & 1.0 & 402 & 2.4 & 13 & -1.10\\
51 & B1727-47 & 342.57 & -7.67& 123.33 & 8.8$\pm$0.2& 71.31 & 0.9 & 125 & 2.8 & 3 & -1.40\\
52 & {\bf B1729-41} & 347.98 & -4.46& 195.30 & 26$\pm$4& 24.53 & 1.2 & 252 & 3.9 & 28 & -1.27\\
53 & B1737-30$^{[1]}$ $\ast$ & 358.29 & 0.24& 152.15 & 117$\pm$6& 200.28 & 1.0 & 196 & 2.7 & 11 & -0.44\\
54 & B1738-08 $\dagger$ & 16.96 & 11.30& 74.90 & 2.4$\pm$0.2& 105.52 & 1.1 & 481 & 2.2 & 0.3 & -1.67\\
55 & B1737-39 & 350.56 & -4.75& 158.50 & 17.2$\pm$0.4& 56.03 & 1.0 & 253 & 3.2 & 5.3 & -1.27\\
56 & B1740-13 & 12.70 & 8.21& 116.30 & 0.39$\pm$0.05& 19.35 & 1.0 & 374 & 2.9 & 1.5 & -2.56\\
57 & B1740-31 & 357.29 & -1.15& 193.05 & 296$\pm$24& 244.03 & 1.0 & 471 & 3.3 & 29 & -0.27\\
58 & {\bf B1742-30} $\dagger$ & 358.55 & -0.963& 88.37 & 3.7$\pm$0.2& 35.88 & 1.1 & 253 & 1.9 & 2.7 & -1.41\\
59 & {\bf J1750-3503} & 355.31 & -4.08& 189.35 & 51$\pm$7& 101.54 & 1.1 & 252 & 3.9 & 48 & -1.03\\
60 & {\bf B1756-22} & 7.47 & 0.81& 177.16 & 9.0$\pm$0.3& 19.72 & 1.0 & 441 & 3.6 & 17 & -1.59\\
61 & B1758-03 & 23.60 & 9.26& 120.37 & 0.6$\pm$0.1& 29.66 & 1.0 & 432 & 3.6 & 0.7 & -2.58\\
62 & B1804-12 $\dagger$ & 17.14 & 4.42& 122.41 & 2.0$\pm$0.1& 40.83 & 1.0 & 125 & 2.8 & 2.2 & -1.93\\
63 & B1804-08 $\dagger$ & 20.06 & 5.59& 112.38 & 4.3$\pm$0.1& 22.47 & 1.0 & 150 & 2.7 & 4.4 & -1.74\\
64 & {\bf B1804-27} & 3.84 & -3.26& 312.98 & 37$\pm$2& 90.54 & 1.3 & 125 & 5.0 & 31 & -1.35\\
65 & J1808-0813 & 20.63 & 5.75& 151.27 & 12$\pm$1& 43.80 & 0.8 & 396 & 3.8 & 2 & -1.54\\
66 & B1813-26 $\dagger$ & 5.22 & -4.91& 128.12 & 7.5$\pm$0.3& 58.51 & 0.9 & 149 & 3.1 & 2.8 & -1.92\\
67 & J1817-3837 & 354.68 & -10.41& 102.85 & 4$\pm$1& 13.52 & 1.2 & 252 & 2.5 & 1 & -1.55\\
68 & B1818-04 $\dagger$ & 25.46 & 4.73& 84.44 & 2.74$\pm$0.02& 48.15 & 2.7 & 469 & 1.9 & 3.7 & -1.53\\
69 & B1819-22 $\dagger$ & 9.35 & -4.37& 121.20 & 14.5$\pm$0.2& 161.05 & 1.0 & 253 & 3.0 & 2.7 & -1.27\\
70 & {\bf J1823-0154} & 28.08 & 5.26& 135.87 & 5.9$\pm$0.3& 20.95 & 1.2 & 396 & 3.6 & 8.2 & -1.75\\
71 & B1821-19 & 12.28 & -3.11& 224.65 & 56$\pm$3& 77.01 & 1.0 & 173 & 4.7 & 25 & -1.15\\
72 & B1823-11 & 19.80 & 0.29& 320.58 & 163$\pm$13& 149.52 & 1.0 & 444 & 4.6 & 133 & -0.73\\
73 & {\bf B1826-17} & 14.60 & -3.42& 217.11 & 92$\pm$4& 132.43 & 1.1 & 266 & 4.7 & 90 & -0.96\\
74 & B1829-08 $\dagger$ & 23.27 & 0.29& 300.87 & 52$\pm$5& 54.98 & 0.9 & 361 & 4.9 & 239 & -1.20\\
75 & {\bf B1831-03} & 27.66 & 2.27& 234.54 & 50$\pm$1& 181.46 & 1.0 & 289 & 5.1 & 29 & -1.26\\
76 & J1835-1106 & 21.22 & -1.51& 132.68 & 6.8$\pm$0.4& 12.06 & 1.0 & 395 & 2.8 & 4.0 & -1.51\\
77 & B1834-10 $\ast$ & 22.26 & -1.42& 316.98 & 410$\pm$184& 124.17 & 0.7 & 297 & 4.5 & 96 & -0.38\\
78 & B1839-04 $\dagger$ & 28.35 & 0.17& 195.98 & 215$\pm$11& 618.78 & 1.1 & 331 & 4.7 & 2.4 & -0.71\\
79 & B1841-04 & 28.10 & -0.55& 123.16 & 43$\pm$2& 74.55 & 1.0 & 449 & 2.9 & 3.2 & -0.87\\
80 & B1844-04 & 28.88 & -0.94& 141.98 & 33$\pm$1& 91.29 & 0.9 & 474 & 3.3 & 6 & -1.05\\
81 & {\bf B1845-01} & 31.34 & 0.04& 159.53 & 115$\pm$5& 140.48 & 1.3 & 456 & 4.0 & 106 & -0.75\\
82 & {\bf J1848-1414} & 19.90 & -5.84& 134.47 & 8$\pm$1& 8.19 & 1.2 & 397 & 3.6 & 33 & -1.84\\
83 & B1846-06 & 26.77 & -2.50& 148.17 & 24$\pm$1& 78.46 & 1.3 & 478 & 3.4 & 4 & -1.20\\
84 & B1852+10 & 42.89 & 4.22& 207.20 & 36$\pm$2& 44.78 & 1.3 & 252 & 6.9 & 2 & -1.50\\
85 & B1851-14 & 20.46 & -7.21& 130.40 & 0.8$\pm$0.1& 51.70 & 1.0 & 485 & 3.7 & 1.2 & -2.49\\
86 & B1859+01 & 35.82 & -1.37& 105.39 & 0.73$\pm$0.04& 8.62 & 1.0 & 298 & 2.8 & 0.45 & -2.30\\
87 & B1859+03 $\ast \dagger$  & 37.21 & -0.64& 402.08 & 177$\pm$6& 390.28 & 0.8 & 170 & 7.3 & 182 & -1.08\\
88 & B1859+07 & 40.57 & 1.06& 252.81 & 23$\pm$5& 27.93 & 1.0 & 434 & 5.5 & 80 & -1.58\\
89 & {\bf B1900+05} & 39.50 & 0.21& 177.49 & 29$\pm$3& 30.43 & 1.1 & 394 & 4.7 & 29 & -1.37\\
90 & B1900+06 & 39.81 & 0.34& 502.90 & 197$\pm$20& 265.14 & 1.5 & 123 & 8.4 & 1085 & -1.16\\
91 & B1900+01 & 35.73 & -1.96& 245.17 & 28.2$\pm$0.3& 81.19 & 1.0 & 1021 & 3.3 & 43 & -1.12\\
92 & {\bf B1900-06} & 28.48 & -5.68& 195.61 & 20.7$\pm$0.3& 92.98 & 0.9 & 365 & 5.4 & 20 & -1.61\\
93 & {\bf J1904+0004} & 34.45 & -2.81& 233.61 & 15$\pm$2& 25.10 & 1.2 & 112 & 5.7 & 11 & -1.78\\
94 & J1904-1224 & 23.29 & -8.49& 118.23 & 5.6$\pm$0.2& 26.28 & 0.9 & 397 & 3.4 & 1 & -1.69\\
95 & {\bf B1902-01} & 33.69 & -3.55& 229.13 & 7.4$\pm$0.4& 25.67 & 0.5 & 473 & 6.0 & 10 & -2.07\\
96 & J1908+0500 & 39.29 & -1.40& 201.42 & 9$\pm$2& 6.82 & 1.0 & 252 & 5.2 & 25 & -1.89\\
97 & B1907+02 $\dagger$ & 37.61 & -2.71& 171.73 & 0.4$\pm$0.1& 37.20 & 0.9 & 476 & 4.9 & 3.9 & -2.31\\
98 & B1907+10 $\dagger$ & 44.83 & 0.99& 149.98 & 1.35$\pm$0.02& 11.36 & 0.8 & 1021 & 4.2 & 3.2 & -2.40\\
99 & B1907-03 $\dagger$ & 32.28 & -5.68& 205.53 & 2.69$\pm$0.04& 24.64 & 0.9 & 1021 & 6.1 & 1.4 & -2.45\\
100 & J1910+0714 & 41.52 & -0.87& 124.06 & 4$\pm$1& 42.38 & 0.7 & 509 & 4.1 & 4.2 & -2.03\\
101 & B1907+12 & 46.22 & 1.59& 258.64 & 52$\pm$5& 31.60 & 1.0 & 363 & 7.4 & 22 & -1.53\\
102 & B1911-04 $\dagger$ & 31.31 & -7.12& 89.39 & 0.19$\pm$0.01& 29.23 & 0.6 & 1020 & 2.8 & 1.65 & -2.11\\
103 & B1911+13 $\dagger$ & 47.88 & 1.59& 145.05 & 0.5$\pm$0.1& 25.75 & 1.0 & 402 & 5.1 & 6.4 & -2.92\\
104 & B1914+13 & 47.58 & 0.45& 237.01 & 6$\pm$1& 10.68 & 1.0 & 392 & 5.0 & 11 & -1.99\\
105 & B1915+13 & 48.26 & 0.62& 94.54 & 0.36$\pm$0.01& 13.33 & 1.2 & 216 & 4.0 & 3.05 & -2.84\\
106 & B1918+19 & 53.87 & 2.67& 153.85 & 9.4$\pm$0.3& 86.43 & 1.0 & 282 & 5.6 & 0.62 & -1.64\\
107 & B1920+21 $\dagger$ & 55.28 & 2.94& 217.09 & 2.3$\pm$0.1& 44.48 & 1.1 & 409 & 7.6 & 4.6 & -2.69\\
108 & B1922+20 & 55.02 & 2.33& 213.00 & 9$\pm$4& 3.98 & 0.8 & 235 & 7.3 & 1.3 & -2.15\\
109 & B1923+04 & 40.98 & -5.67& 102.24 & 2.2$\pm$0.1& 26.67 & 0.9 & 440 & 3.8 & 0.2 & -2.15\\
110 & B1924+14 $\dagger$ & 49.92 & -1.04& 211.41 & 19$\pm$1& 36.55 & 1.0 & 432 & 6.7 & 13 & -1.82\\
111 & B1924+16 & 51.86 & 0.06& 176.88 & 4.6$\pm$0.3& 18.66 & 1.0 & 370 & 5.8 & 7 & -2.22\\
112 & B1929+20 & 55.58 & 0.64& 211.15 & 19$\pm$1& 28.60 & 1.1 & 419 & 6.9 & 18 & -1.84\\
113 & B1930+22 & 57.36 & 1.55& 219.20 & 0.65$\pm$0.03& 5.58 & 1.1 & 230 & 7.5 & 1.4 & -3.13\\
114 & B1933+16 & 52.44 & -2.09& 158.52 & 3.21$\pm$0.02& 28.62 & 1.0 & 373 & 5.6 & 0.9 & -2.32\\
115 & B1944+22 & 59.30 & -1.07& 140.00 & 12$\pm$1& 31.28 & 1.2 & 253 & 5.4 & 1 & -1.81\\
116 & B1946+35 & 70.70 & 5.05& 129.07 & 35.7$\pm$0.3& 138.63 & 0.8 & 473 & 5.8 & 28 & -1.48\\
117 & B2000+32 & 69.26 & 0.88& 142.21 & 9$\pm$1& 25.07 & 0.9 & 414 & 5.7 & 0.6 & -1.96\\
118 & B2002+31 & 69.01 & 0.02& 234.82 & 9.0$\pm$0.2& 52.67 & 1.2 & 478 & 7.5 & 10 & -2.18\\
119 & {\bf B2011+38} & 75.93 & 2.48& 238.22 & 22$\pm$2& 34.99 & 1.0 & 325 & 8.4 & 16 & -1.94\\
120 & B2027+37 & 76.89 & -0.73& 190.66 & 11.6$\pm$0.3& 53.89 & 1.1 & 426 & 6.3 & 3 & -1.95\\
121 & B2053+36 & 79.13 & -5.59& 97.31 & 5.5$\pm$0.3& 16.56 & 0.9 & 385 & 4.6 & 5.8 & -1.97\\
122 & B2106+44 & 86.91 & -2.01& 139.83 & 16$\pm$1& 33.32 & 1.1 & 494 & 5.0 & 1.4 & -1.64\\
123 & B2111+46 $\dagger$ & 89.00 & -1.27& 141.26 & 1.8$\pm$0.1& 234.48 & 0.7 & 235 & 4.5 & 1 & -2.36\\
124 & B2319+60 $\dagger$ & 112.10 & -0.57& 94.59 & 11$\pm$1& 152.90 & 1.2 & 425 & 3.0 & 1 & -1.40\\
\enddata
\vspace{-0.6cm}
\noindent \tablecomments{[1] Since it is a glitching pulsar, the period provided by TEMPO2 predictors can have an offset.
Hence the period of the pulsar is found from several observations with ORT. Since this solution can also have unknown errors 
in period, the $\tau_{sc}$ obtained can have larger uncertainty.\\$\ast$ --- denote pulsars for which the pulse 
broadening function PBF2 from LKMLL01 was also fitted as discussed in Section~\ref{analp}. The fitted results and plots are 
provided in Table 2 and Figure 63 of the supplementary material.
\\$\dagger$ --- denote pulsars with multicomponent profiles, where only one of the components is fitted to get an estimate 
of the $\tau_{sc}$, due to the effects of profile evolution as discussed in Section~\ref{analp}.}
\label{tab1}
\end{deluxetable}

\end{document}